\documentclass[runningheads]{llncs}
\usepackage[T1]{fontenc}
\usepackage{graphicx}
\usepackage{amsfonts}   % provides \mathbb
\usepackage{amssymb}    % also provides \mathbb and more
\usepackage{amsmath}
\usepackage{xcolor}
\usepackage{mathrsfs} 
\newtheorem{examples}{Example}

\usepackage{array}      % for \arraybackslash and advanced column definitions
\usepackage{booktabs}   % for \toprule, \midrule, \bottomrule
\usepackage{multirow}   % for multi-row cells

\usepackage{caption}    % for better table captions
\usepackage{graphicx}   % if you use images
\usepackage{enumerate,enumitem}

\usepackage{mathrsfs}

\usepackage{array}

\usepackage{booktabs}

\usepackage{multirow}

\usepackage{algorithm}
\usepackage{algpseudocode}

\usepackage{subcaption}   % preferred over old 'subfigure' package

\begin{document}
\title{Algebraic Characterization of Reversible First Degree Cellular Automata over $\mathbb{Z}_d$}
% \author{}
% \institute{}
% \vspace{-2em}
\titlerunning{Algebraic Characterization of Reversible FDCAs over $\mathbb{Z}_d$}
\author{Baby C. J. \inst{1} \and Kamalika Bhattacharjee\inst{2}}
\institute{Department of Computer Science and Engineering, National Institute of Technology, Tiruchirappalli, Tamil Nadu 620015, India \\
 \and
Department of Information Technology, Indian Institute of Engineering Science and Technology, Shibpur, West Bengal 711103, India \\
\email{\{babycj1120, kamalika.it\}@gmail.com}}

\maketitle              % typeset the header of the contribution
\begin{abstract}
There exists algorithms to detect reversibility of cellular automaton (CA) for both finite and infinite lattices taking quadratic time. But, can we identify a $d$-state CA rule in constant time that is always reversible for every lattice size $n\in \mathbb{N}$? To address this issue, this paper explores the reversibility properties of a subset of one-dimensional, $3$-neighborhood, $d$-state finite cellular automata (CAs), known as the first degree cellular automata (FDCAs) for any number of cells $(n\in \mathbb{N})$ under the null boundary condition. {In a first degree cellular automaton (FDCA), the local rule is defined using eight parameters. To ensure that the global transition function of $d$-state FDCA is reversible for any number of cells $(n\in \mathbb{N})$, it is necessary and sufficient to verify only three algebraic conditions among the parameter values.
% These conditions are identified, proven, and validated in this study. 
Based on these conditions, for any given $d$, one can synthesize all reversible FDCAs rules. Similarly, given a FDCA rule, one can check these conditions to decide its reversibility in constant time.}

\end{abstract}

 % First-Degree Cellular Automata \sep Reversibility

%
%
%
\section{Introduction}
Reversibility in Cellular Automata (CAs) refers to the property where every configuration has exactly one predecessor. This implies that the configurations of a reversible cellular automaton follow a one-to-one relationship. The study of reversibility has been ongoing for many years and finds applications in various fields \cite{chaudhuri1997additive,bhattacharjee2020survey}, including pattern recognition, clustering, pseudo-random number generation, data classification, cryptography, and more.

In classical literature, the study of the reversibility of CAs is primarily done on infinite lattices \cite{amoroso1972decision,sutner1991bruijn}, although there are some works that address the reversibility of CAs on finite lattices \cite{bhattacharjee2016reversibility,bhattacharjee2019finite,das2006theory,das2009characterization} as well. A CA is considered to be reversible if its global transition function is bijective, meaning it is both injective (one-to-one) and surjective (onto). In the classical work by Amorosso and Patt \cite{amoroso1972decision}, the reversibility of CAs, defined over infinite lattice sizes, is determined using specific algorithms to test the properties of surjectivity and injectivity. Whereas, in Sutner's paper \cite{sutner1991bruijn}, the reversibility of the same kind of CAs is analyzed through various properties of de Bruijn graphs. Later literature explores the reversibility of CAs with particular number of cells $(n)$, specifically in finite lattices \cite{das2006theory,bhattacharjee2016reversibility}. 
Other studies examine the relationship between the reversibility of CAs in finite and infinite latices, classifying the reversibility into three classes: \emph{reversible, strictly irreversible}, and \emph{semi-reversible} \cite{bhattacharjee2019finite}. A CA with rule $R$ is $reversible$ if its global transition function is bijective on the set of all configurations for all number of cells $(n \in \mathbb{N})$. In contrast, a CA is \emph{strictly irreversible} if the global transition function is not bijective for every number of cells $(n \in \mathbb{N})$. A CA with rule $R$ is termed as \emph{semi-reversible} if the global transition function is bijective on the set of all configurations for some specific number of cells $(n)$. These three classes of CAs are based on reversibility defined in the finite lattice under periodic boundary condition. When we examine an infinite lattice, the class of $reversible$ CA rules remains the same, but the classes \emph{strictly irreversible} and \emph{semi-reversible} are considered as \emph{irreversible} in the context of infinite lattice. Hereafter, unless explicitly mentioned, by ``a reversible CA'', we mean the CA is reversible for all $n \in \mathbb{N}$.

{In all existing works on the reversibility of cellular automata, there is no way to synthesize the set of all reversible rules without testing its reversibility. Even the best algorithm for deciding the reversibility of a CA under periodic configurations (for all $n \in \mathbb{N}$) takes at least quadratic time of the number of nodes of the de Bruijn graph as complexity \cite{sutner1991bruijn}. But, there is no such algorithm developed for reversibility over all $n$ under null boundary condition. In addition, there is no synthesis technique that guaranties the generation of all possible reversible CAs for any $d$. If it were possible to identify whether a CA rule is reversible for all cells simply by examining the rule itself, the analysis and application of reversible CAs would be much easier. However, for $d$-state $m$-neighborhood CAs, in general, such an analysis is extremely difficult. Therefore, we focus on a subset of $3$-neighborhood $d$-state CAs, called first degree CAs (FDCAs) \cite{bhattacharjee2022first} under null boundary condition.}  In the case of the FDCA, the local rule is defined using eight parameters over $\mathbb{Z}_d$. This study shows that an FDCA is reversible for any number of cells $(n \in \mathbb{N})$ if and only if these parameters satisfy three specific algebraic conditions. 
 These three conditions are identified, proven, and verified in this work (Section~\ref{sec:Reversibility of First Degree CAs}). For any $d$-state FDCA, the list of valid FDCA parameters satisfying these conditions constitutes the set of reversible FDCA rules under null boundary conditions (Section~\ref{sec: Case Study of Reversible FDCAs}). Using these conditions, a constant time algorithm is also proposed that scans the eight FDCA parameters of any given $d$-state FDCA rule and decides its reversibility (Section~\ref{sec: Reversibility Testing}).

\vspace{-.3 cm}
\section{First Degree Cellular Automata}
In this work, we consider one-dimensional 3-neighborhood finite CAs under the null boundary condition, where each cell changes its state depending on its current state and the states of its left and right neighbors. 
% The cells of the CA form a lattice $\mathscr{L}=\mathbb{Z}/n\mathbb{Z}$, where $n$ is the number of cells of the CA. 
The cells are numbered from $0$ to $n-1$, each cell of such an $n$-cell CA uses a set of states $S=\{0,1,2,\cdots,d-1\}$. The next state of each cell can be determined by a local rule $R: S^3 \to S$. A configuration $c:\mathscr{L} \to S$ is the states of all the cells at any time instant.
\begin{definition}
    A finite CA of lattice size $n$ with global transition function $G_n : C_n \rightarrow C_n$ is reversible, if for each $y \in C_n$, where $C_n$ is the set of all configurations of length $n$, there exists exactly one $x \in C_n$, such that, $G_n(x)=y$. That is, the CA is reversible if and only if $G_n$ is a bijection. Otherwise, the CA is irreversible for that $n$ \cite{DBLP:journals/corr/abs-1911-03609}.
\end{definition}

A CA is called \emph{reversible}, if it is reversible for each lattice size $n \in \mathbb{N}$. There exists algorithms to decide reversibility of finite CAs for a specific $n$ under null boundary condition \cite{das2006theory,mukherjee2021reversible}. So, deciding reversibility for all $n$ will involve running a CA with such an algorithm for all $n \in \mathbb{N}$ which is an impossible task. Also, as discussed, there is no efficient technique to synthesize the set of all reversible rules for any $d$-state CA. To address this, we consider a subset of $3$-neighborhood $d$-state CAs, called the first-degree cellular automata (FDCAs).
% \vspace{-.6 cm}
 Introduced in Ref.\cite{bhattacharjee2022first}, a first-degree cellular automaton is a very effective representation for a subset of CAs where the number of states per cell can be large but the neighborhood is restricted to only three. Formally, a first-degree rule is defined as the following: 
\begin{definition}\label{Def:FDCA}
    A cellular automaton rule $R:S^3\rightarrow S$ is of first degree, if the rule $R: S^3 \rightarrow S$ can be represented in the following form: $R(x,y,z)=c_0xyz + c_1xy + c_2xz + c_3yz + c_4x + c_5y + c_6z + c_7\pmod d$.
Here, $S$ is the finite state set, $d$ is the number of states of CA, and $x,y,z$ are the states of the three neighbors where $x,y,z \in S$. The constants $c_i \in \mathbb{Z}_d$  \cite{bhattacharjee2022first,bhattacharjee2023study}. 
\end{definition}

\vspace{-.8 cm}
\begin{figure}[h]
    \centering
    \includegraphics[width=.8\linewidth]{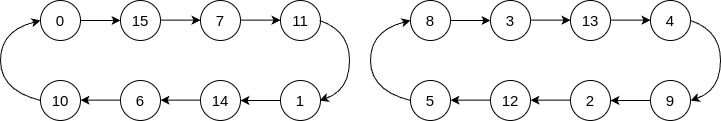}
    \caption{Transition diagram of the 4-cell 2-state FDCA with parameters $\langle0,0,0,0,1,1,0,1\rangle$}
    \label{fig:transition}
\end{figure}
\vspace{-.6 cm}
{Any first-degree cellular automaton can be represented using the eight coefficients of its local rule,
$\langle c_0,\, c_1,\, c_2,\, c_3,\, c_4,\, c_5,\, c_6,\, c_7 \rangle$. These coefficients are called the \emph{parameters} of FDCA rule. For example, Fig \ref{fig:transition} represents the transition diagram of the 4-cell 2-state FDCA with parameters $\langle0,0,0,0,1,1,0,1\rangle$.} Here, configuration 1010 (10) is reachable from configuration 0000 (0), whereas configuration 1000 (8) is not reachable from 0000 (0). However every configuration has a unique predecessor, there are no non-reachable configurations. Hence, this FDCA is reversible for $n=4$.

\vspace{-.3 cm}

\section{Reversibility of First Degree CAs}
\label{sec:Reversibility of First Degree CAs}
{In this section, we present the conditions that must be satisfied by the parameters of the FDCAs to become reversible. We then formulate the corresponding theories associated with each condition.} {Let us begin with an example FDCA rule over the state set $\mathbb{Z}_d$}.

\begin{examples}
Consider an FDCA rule over the state set $S=\mathbb{Z}_8$ (i.e., $d = 8$), with parameters $\langle c_0,~c_1,$ $~c_2,~c_3,$ $~c_4,~c_5,$ $~c_6,~c_7 \rangle$ $ = \langle 2, 4, 0, 6, 3, 3, 2, 1\rangle$. This rule is reversible for each lattice size $n \in \mathbb{N}$. {We now examine the algebraic conditions satisfied by this rule. Here, the greatest common divisor between $c_5$ and $d$ = $\gcd(c_5, d) = \gcd(3, 8) = 1$, indicating that $c_5$ is relatively prime to $d$. Moreover, the coefficient $c_7=1$, which belongs to the state set $\mathbb{Z}_8$.
Here, the parameters $c_0,~c_1,~c_2,$ and $c_3$ take the values $2,~4,~0,$ and $6$, respectively. For $d = 8$, we have $\mathrm{rad}(8) = 2$, where $\mathrm{rad}(d)$ denotes the product of the distinct prime factors of $d$, and hence the corresponding set comprises all even numbers. Therefore, the coefficients $c_0,~c_1,~c_2,$ and $c_3$ are all multiples of $\mathrm{rad}(d)$. That means each of these values belongs to the set $\{a \mid a \equiv 0 \pmod{\mathrm{rad}(d)}\}$.  For $c_4 = 3$ and $c_6 = 2$, the product $c_4 \cdot c_6=6 $ is also multiple of $\mathrm{rad}(d)$, which means the coefficient $c_4$ and $c_6$ satisfies the condition $c_4 \cdot c_6 \equiv 0 \pmod{\mathrm{rad}(d)}$.}
\end{examples}

{All reversible FDCA rules, as demonstrated in the above example, satisfy certain algebraic conditions. Specifically, only those rules whose parameters satisfy the following three conditions are reversible for all $n \in \mathbb{N}$.}

\begin{enumerate}
    \item $\gcd(c_5,d)=1$ where $c_5\ne0$ and $c_7\in \mathbb{Z}_d$
    \item $c_0,c_1,c_2,c_3\in\{a|$ $a \equiv 0 \pmod{\mathrm{rad}(d)}$, where $\mathrm{rad}(d)$ is the product of distinct primes dividing $d$. That is $\mathrm{rad}(d)= \displaystyle\prod_{p \mid d} p\}$
    \item $c_4 \cdot c_6 \equiv 0 \pmod{\mathrm{rad}(d)}$,
where $\mathrm{rad}(d)$ is the product of the distinct primes dividing $d$.

\end{enumerate}
Any rule that fails to follow these conditions is irreversible rule for some or all number of cells, indicating they will be in either \emph{semi-reversible} or \emph{strictly-irreversible} categories. We are interested to identify the properties of the reversible rules. {The following lemmas correspond to these three conditions:}
\begin{lemma}
\label{lemma:c5}
    If an FDCA rule is reversible for all number of cells $n\in \mathbb{N}$, then $\gcd(c_5,d)=1$ where $c_5\ne0$ and $c_7\in \mathbb{Z}_d$.
\end{lemma}
% \begin{proof}
%    Consider, a FDCA rule is reversible for all $n\in \mathbb{N}$. That means, the CA is also reversible for cell length $n=1$. Now, to prove by contradiction, suppose that $\gcd(c_5, d) \ne 1$  where $c_5\ne0$. 
%     Then, for any two configurations $u$ and $v$ be of length $1$, where $u\ne v$, the next configuration is:
%     \vspace{-.2 cm}
% \begin{align*}
% R(0,u,0) &= c_5 u + c_7 \pmod{d} && \text{(Under null boundary condition)} \\
%          &= c_7 \pmod{d} && \text{(As $\gcd(c_5,d) \ne 1$, $c_5$ is a multiple of $d$)} \\
% R(0,v,0) &= c_5 v + c_7 \pmod{d} && \text{(Under null boundary condition)} \\
%          &= c_7 \pmod{d} && \text{(As $\gcd(c_5,d) \ne 1$, $c_5$ is a multiple of $d$)}
% \end{align*}  
%    This implies the FDCA is irreversible since two different configurations map to the same configuration. Now, consider the FDCA with $n$ cells, where $n \in \mathbb{N}$. Take two configurations $0^{n-1} u ~\text{and}~0^{n-1} v, ~ u \ne v$. By the same reasoning, if $\gcd(c_5, d) \ne 1$, these two configurations 
% will map to the same next configuration. In such cases, $c_5$ and $d$ share a common factor, which makes the FDCA irreversible. Hence, we reach a contradiction. Therefore, if an FDCA is reversible, then $\gcd(c_5, d) = 1$ and $c_5 \ne 0$ has to hold.
%  Here, $c_7$ is an additive constant that does not affect reversibility; that means $c_7\in \mathbb{Z}_d$.
% \end{proof}
\begin{examples}
\setlength{\abovedisplayskip}{4pt}
\setlength{\belowdisplayskip}{4pt}
\setlength{\abovedisplayshortskip}{2pt}
\setlength{\belowdisplayshortskip}{2pt}

Consider a $d$-state FDCA with $d = 6$ under null boundary conditions. Let the parameters be
$\langle c_0,c_1,c_2,c_3,c_4,c_5,c_6,c_7\rangle
= \langle0,0,0,0,0,3,0,1\rangle$ so that the local rule becomes $R(x,y,z) = 3y + 1 \pmod{6}.$ Take two distinct configurations \((u,v,w) = (0,0,2)\) and \((r,s,t) = (0,0,4)\). Under null boundary, the next configurations are:\\ For $(u,v,w)=(0,0,2)$:
\begin{align*}
u'_0 &= R(0,u,v) = R(0,0,0) = 3\cdot0 + 1 \equiv 1 \pmod{6},\\
u'_1 &= R(u,v,w) = R(0,0,2) = 3\cdot0 + 1 \equiv 1 \pmod{6},\\
u'_2 &= R(v,w,0) = R(0,2,0) = 3\cdot2 + 1 = 7 \equiv 1 \pmod{6},
\end{align*}
% hence $(0,0,2) \mapsto (1,1,1)$.

\noindent For $(r,s,t)=(0,0,4)$:
\begin{align*}
r'_0 &= R(0,r,s) = R(0,0,0) = 3\cdot0 + 1 \equiv 1 \pmod{6},\\
r'_1 &= R(r,s,t) = R(0,0,4) = 3\cdot0 + 1 \equiv 1 \pmod{6},\\
r'_2 &= R(s,t,0) = R(0,4,0) = 3\cdot4 + 1 = 13 \equiv 1 \pmod{6},
\end{align*}
% hence $(0,0,4) \mapsto (1,1,1)$.

Therefore, the two distinct input configurations evolve to the same configuration: $(0,0,2)\mapsto(1,1,1),~ (0,0,4)\mapsto(1,1,1)$. Thus, the FDCA is \emph{irreversible}. Here $c_5 = 3$ and $\gcd(c_5,6) = 3 \neq 1$, which violates the condition for reversibility.
\end{examples}
This lemma indicates, if this condition is violated, the CA will be \emph{strictly irreversible}, that is, irreversible for all $n \in \mathbb{N}$.

\begin{lemma}
\label{lemma:c0,1,2,3}
    If an FDCA rule is reversible for any all $n\in \mathbb{N}$, then $c_0=c_1=c_2=c_3\equiv 0 \pmod{\mathrm{rad}(d)}$, where $\mathrm{rad}(d)$ is the product of distinct primes dividing $d$. 
\end{lemma}

\begin{examples}
\setlength{\abovedisplayskip}{4pt}
\setlength{\belowdisplayskip}{4pt}
\setlength{\abovedisplayshortskip}{2pt}
\setlength{\belowdisplayshortskip}{2pt}

Consider a $d$-state FDCA with $d=8$ under null boundary conditions. Let the parameters be
$\langle c_0,c_1,c_2,c_3,c_4,c_5,c_6,c_7\rangle
=\langle0,1,0,0,5,1,2,0\rangle$
so that $R(x,y,z)\equiv xy+5x+y+2z \pmod{8}$ $\equiv y(x+1)+5x+2z \pmod{8}$. Take two distinct configurations \((u,v)=(7,0)\) and \((r,s)=(7,4)\) for $n=2$. Under null boundary, the next configurations are:
\begin{align*}
R(0,u,v)&=R(0,7,0)=0+0+7+0=7\equiv7\pmod{8},\\
R(u,v,0)&=R(7,0,0)=0+35+0+0=35\equiv3\pmod{8},\\
R(0,r,s)&=R(0,7,4)=0+0+7+8=15\equiv7\pmod{8},\\
R(r,s,0)&=R(7,4,0)=28+35+4+0=67\equiv3\pmod{8}.
\end{align*}

Hence, $(u,v)\mapsto(7,3),~ (r,s)\mapsto(7,3).$ Both distinct inputs evolve to the same next configuration \((7,3)\). Therefore, for this parameter set, $c_1\not\equiv0\pmod{\mathrm{rad}(d)}$. Since $\mathrm{rad}(8)=2$ and $c_1=1$, we have $c_1\not\equiv0\pmod{2}$, and the FDCA is \emph{irreversible}.
\end{examples}

\begin{lemma}
\label{lemma:c4,6}
If the FDCA rule is reversible for any number of cells $n\in \mathbb{N}$, then $c_4 \cdot c_6 \equiv 0 \pmod{\mathrm{rad}(d)}$,
where $\mathrm{rad}(d)$ is the product of the distinct primes dividing $d$.
\end{lemma}

\begin{examples}
Consider a $d$-state FDCA with $d=12$ under null boundary conditions. Let the parameters be $\langle c_0,c_1,c_2,c_3,c_4,c_5,c_6,c_7\rangle=\langle 0,0,0,0,5,1,1,0\rangle.$ Consider two distinct configurations $(u,v)=(6,3)$ and $(v,u)=(3,6)$. The next configurations are:
\begin{align*}
R(0,u,v)&=5(0)+1(u)+1(v)\equiv u+v\equiv9\pmod{12},\\
R(u,v,0)&=5(u)+1(v)+0\equiv5u+v\equiv5(6)+3=33\equiv9\pmod{12},\\
R(0,v,u)&=5(0)+1(v)+1(u)\equiv v+u\equiv9\pmod{12},\\
R(v,u,0)&=5(v)+1(u)+0\equiv5v+u\equiv5(3)+6=21\equiv9\pmod{12}.
\end{align*}
In both cases, the resulting configuration is identical $(u,v)\mapsto(9,9),~(v,u)\mapsto(9,9$).
Hence, two distinct input configurations lead to the same next configuration. Here, $c_4 = 5$ and $c_6 = 1$.
% The product $c_4 \cdot c_6 = 5$, which is not a multiple of $\mathrm{rad}(12)$. 
Since $\mathrm{rad}(12) = 2 \times 3 = 6$ and $c_4 \cdot c_6 \not\equiv 0 \pmod{\mathrm{rad}(12)}$, the FDCA is \emph{irreversible} for these parameter values.
\end{examples}
The following theorem gives the necessary and sufficient conditions for reversibility in FDCAs.
% \smallskip
% \noindent\textbf{Case 2.}
% Now consider another $d$-state FDCA with parameters
% \[
% \langle c_0, c_1, c_2, c_3, c_4, c_5, c_6, c_7\rangle
% = \langle 0,0,0,0,1,1,6,0\rangle,
% \]
%  under null boundary conditions.
% For the same pair of distinct configurations $(u,v)=(6,3)$ and $(v,u)=(3,6)$,
% the next states are
% \begin{align*}
% R(0,u,v) &= 0\cdot c_4 + 6\cdot c_5 + 3\cdot c_6 
% \equiv 6 + 18 \equiv 24 \equiv 0 \pmod{12},\\[4pt]
% R(u,v,0) &= 6\cdot c_4 + 3\cdot c_5 + 0\cdot c_6 
% \equiv 6 + 3 \equiv 9 \pmod{12},\\[4pt]
% R(0,v,u) &= 0\cdot c_4 + 3\cdot c_5 + 6\cdot c_6 
% \equiv 3 + 36 \equiv 39 \equiv 3 \pmod{12},\\[4pt]
% R(v,u,0) &= 3\cdot c_4 + 6\cdot c_5 + 0\cdot c_6 
% \equiv 3 + 6 \equiv 9 \pmod{12}.
% \end{align*}
% Thus, the two configurations evolve as
% \[
% (u,v) \mapsto (0,9), \qquad (v,u) \mapsto (3,9),
% \]
% which are distinct.

% In case 2, $\mathrm{rad}(12)=6$ and $c_4 c_6 = 1\times6 = 6 \equiv 0 \pmod{6}$.
% Therefore, the condition 
% $c_4 c_6 \equiv 0 \pmod{\mathrm{rad}(d)}$
% is satisfied.
\begin{theorem}
\label{theorem: irrversible}
\vspace{-.6em}
The rules of $d$-state first-degree cellular automata are reversible for all number of cells $(n \in \mathbb{N})$ if and only if the rules follow the three conditions given below.
\begin{enumerate}[noitemsep,topsep=0pt]
    \item $GCD(c_5,d)=1$ and $c_7\in \mathbb{Z}_d$
    \item $c_0,c_1,c_2,c_3\in\{a|$ $a \equiv 0 \pmod{\mathrm{rad}(d)}$, where $\mathrm{rad}(d)$ is the product of distinct primes dividing $d$. That is $\mathrm{rad}(d)= \displaystyle\prod_{p \mid d} p\}$
    \item $c_4 \cdot c_6 \equiv 0 \pmod{\mathrm{rad}(d)}$,
where $\mathrm{rad}(d)$ is the product of the distinct primes dividing $d$.
\end{enumerate}
\end{theorem}

\vspace{-.6 cm}

\begin{table*}[!h]
\centering
\caption{FDCA coefficients satisfying reversibility conditions for various state $d$}
\label{tab:FDCA_rules_compact_final}
\renewcommand{\arraystretch}{1.05}
\setlength{\tabcolsep}{3pt}
\resizebox{0.85\textwidth}{!}{%
\begin{tabular}{
>{\centering\arraybackslash}p{0.3cm}
>{\centering\arraybackslash}p{1.0cm}
>{\centering\arraybackslash}p{0.3cm}
>{\centering\arraybackslash}p{1.0cm}
>{\centering\arraybackslash}p{1.0cm}
>{\centering\arraybackslash}p{1.0cm} |
>{\centering\arraybackslash}p{0.3cm}
>{\centering\arraybackslash}p{1.0cm}
>{\centering\arraybackslash}p{0.3cm}
>{\centering\arraybackslash}p{1.0cm}
>{\centering\arraybackslash}p{1.7cm}
>{\centering\arraybackslash}p{0.8cm}
}
\toprule
\multicolumn{6}{c|}{\textbf{$d$ is Prime }} &
\multicolumn{6}{c}{\textbf{$d$ is Composite }} \\
\cmidrule(lr){1-6} \cmidrule(lr){7-12}
$d$ & $c_0,c_1,$ & $c_4$ & $c_5$ & $c_6$ & $c_7$ &
$d$ & $c_0,c_1,$ & $c_4$ & $c_5$ & $c_6$ & $c_7$ \\
 & $c_2,c_3$ &  &  &  &  &
 & $c_2,c_3$ &  &  &  &  \\
\midrule

% ---- d=2 and d=4 ----
\multirow{2}{*}{2} & \multirow{2}{*}{0}
& 0 & 1 & 0,1 & 0,1 &
\multirow{4}{*}{4} & \multirow{4}{*}{0,2}
& 0 & 1,3 & 0–3 & 0–3 \\
& & 1 & 1 & 0 & 0,1 & & & 1 & 1,3 & 0,2 & 0–3 \\
& &   &   &   &   & & & 2 & 1,3 & 0–3 & 0–3 \\
& &   &   &   &   & & & 3 & 1,3 & 0,2 & 0–3 \\
\midrule

% ---- d=3 and d=6 ----
\multirow{3}{*}{3} & \multirow{3}{*}{0}
& 0 & 1,2 & 0-2 & 0-2 &
\multirow{6}{*}{6} & \multirow{6}{*}{0}
& 0 & 1,5 & 0–5 & 0–5 \\
& & 1 & 1,2 & 0 & 0-2 & & & 1 & 1,5 & 0 & 0–5 \\
& & 2 & 1,2 & 0 & 0-2 & & & 2 & 1,5 & 0,3 & 0–5 \\
& &   &   &   &   & & & 3 & 1,5 & 0,2,4 & 0–5 \\
& &   &   &   &   & & & 4 & 1,5 & 0,3 & 0–5 \\
& &   &   &   &   & & & 5 & 1,5 & 0 & 0–5 \\
\midrule

% ---- d=5 and d=8 ----
\multirow{5}{*}{5} & \multirow{5}{*}{0}
& 0 & 1–4 & 0–4 & 0–4 &
\multirow{8}{*}{8} & \multirow{8}{*}{0,2,4,6}
& 0 & 1,3,5,7 & 0–7 & 0–7 \\
& & 1 & 1–4 & 0 & 0–4 & & & 1 & 1,3,5,7 & 0,2,4,6 & 0–7 \\
& & 2 & 1–4 & 0 & 0–4 & & & 2 & 1,3,5,7 & 0–7 & 0–7 \\
& & 3 & 1–4 & 0 & 0–4 & & & 3 & 1,3,5,7 & 0,2,4,6 & 0–7 \\
& & 4 & 1–4 & 0 & 0–4 & & & 4 & 1,3,5,7 & 0–7 & 0–7 \\
& &   &   &   &   & & & 5 & 1,3,5,7 & 0,2,4,6 & 0–7 \\
& &   &   &   &   & & & 6 & 1,3,5,7 & 0–7 & 0–7 \\
& &   &   &   &   & & & 7 & 1,3,5,7 & 0,2,4,6 & 0–7 \\
\midrule

% ---- d=7 and d=9 ----
\multirow{7}{*}{7} & \multirow{7}{*}{0}
& 0 & 1–6 & 0–6 & 0–6 &
\multirow{9}{*}{9} & \multirow{9}{*}{0,3,6}
& 0 & 1,2,4,5,7,8 & 0–8 & 0–8 \\
& & 1 & 1–6 & 0 & 0–6 & & & 1 & 1,2,4,5,7,8 & 0,3,6 & 0–8 \\
& & 2 & 1–6 & 0 & 0–6 & & & 2 & 1,2,4,5,7,8 & 0,3,6 & 0–8 \\
& & 3 & 1–6 & 0 & 0–6 & & & 3 & 1,2,4,5,7,8 & 0–8 & 0–8 \\
& & 4 & 1–6 & 0 & 0–6 & & & 4 & 1,2,4,5,7,8 & 0,3,6 & 0–8 \\
& & 5 & 1–6 & 0 & 0–6 & & & 5 & 1,2,4,5,7,8 & 0,3,6 & 0–8 \\
& & 6 & 1–6 & 0 & 0–6 & & & 6 & 1,2,4,5,7,8 & 0–8 & 0–8 \\
& &   &   &   &   & & & 7 & 1,2,4,5,7,8 & 0,3,6 & 0–8 \\
& &   &   &   &   & & & 8 & 1,2,4,5,7,8 & 0,3,6 & 0–8 \\
\bottomrule
\end{tabular}}
\end{table*}

\section{Synthesis of Reversible FDCAs}
\label{sec: Case Study of Reversible FDCAs}
In this section, we present a tabular representation of reversible rules for all number of cells corresponding to any state $(d)$, based on the algebraic conditions explained in the previous section. Table \ref{tab:FDCA_rules_compact_final} represents the valid coefficient sets for generating reversible FDCA rules over $\mathbb{Z}_d$. To generate a FDCA reversible rule, we can select one state $d$, then select any one row. The entries in that row specify the allowed values for each coefficient $c_0$ to $c_7$. Choosing any combination of values from these sets ensures that the resulting FDCA rule is reversible.
\vspace{-.3cm}
\subsection{Reversible Rules for $d-$state FDCA where $d$ is prime}
When $d$ is prime, the Theorem~\ref{theorem: irrversible} is simplified. This is because, for a prime $d$, there are no factors. So, the coefficient values $c_0,c_1,c_2,c_3$ are always zero, while the coefficient $c_7$ can take any value in $\mathbb{Z}_d$. {This implies that when $d$ is prime, then all the reversible rules are linear rules.} The coefficient $c_5$ is co-prime to $d$ and $c_4\cdot c_6=0$. 
That is, if $c_4$ is zero, then $c_6$ can take any value from $\mathbb{Z}_d$. However, if $c_4$ is non-zero, then $c_6$ is always zero. This leads to the following corollary.

\begin{corollary}
The rules of a $d$-state FDCA, where $d$ is a prime number, are reversible for all $n \in \mathbb{N}$ if and only if the parameters $c_0 = c_1 = c_2 = c_3 = 0,\quad \text{either } c_4 = 0 \text{ or } c_6 = 0,\quad \text{and } \gcd(c_5, d) = 1.$
\end{corollary}
% \vspace{-1 cm}

% \vspace{-1 cm}

\subsection{Reversible Rules for $d-$state FDCA where $d$ is composite}

When the state of the FDCA rule is composite, then $d$ can be a composite number of the form  $d = p_1^{e_1} p_2^{e_2} \cdots p_k^{e_k}$ where $p_i$s are the prime factors of $d$. Then, the relationship between $c_4$ and $c_6$ is based on the factors of $d$. Specifically, $c_4 \cdot c_6 \equiv 0 \pmod{\mathrm{rad}(d)}$, where $\mathrm{rad}(d)$ is the product of the distinct primes dividing $d$. In this case, the coefficients $c_0, c_1, c_2,$ and $c_3$ can be some non-zero numbers that satisfy the condition $c_0, c_1, c_2, c_3 \in \{ a \,|\, a \equiv 0 \pmod{\mathrm{rad}(d)} \}$. Therefore, for any composite $d$, some non-linear rules also become reversible for all $n\in \mathbb{N}$.
\vspace{-.3 cm}
 \begin{algorithm}[!h]
\caption{Check a FDCA rule is reversible for all number of cells.}\label{algo1}
\label{Algorithm_Rule_analysis}
\begin{algorithmic}[1]
\Require Coefficients $c_0, c_1, c_2, c_3, c_4, c_5, c_6, c_7$ of $d$-state FDCA Rule

\Ensure Report the FDCA rule is reversible or not. 
\State $\mathrm{rad}(d)\longleftarrow$ Product of
distinct primes dividing $d$
\If{$\gcd(c_5,d)\ne 1$ and $c_7\in \mathbb{Z}_d$}
\State Report CA is not reversible and stop.
\EndIf
\For{each of $c_0, c_1, c_2, c_3$} 
\If{$c_i \pmod{\mathrm{rad}(d)} \ne 0$}
\State Report CA is not reversible and stop.
\EndIf
\EndFor
\If{$(c_4 \times c_6 ) \pmod{\mathrm{rad}(d)}\ne0$}
\State Report CA is not reversible and stop.
\EndIf

\If{All the above conditions are satisfied}
\State Report CA is reversible.
\EndIf
\end{algorithmic}

\end{algorithm}
% \vspace{-.5 cm}

\section{Testing Reversibility of FDCAs}
\label{sec: Reversibility Testing}
Instead of generating rules, reversibility can be tested directly using a constant-time algorithm based on coefficient conditions. By examining the coefficients of first-degree cellular automata, we can determine whether a rule is reversible for any number of cells $(n \in N)$. This can be done using Algorithm 1 since only comparison steps result in a complexity of $O(8)$.
\begin{figure}[!h]
    \centering
    
    \begin{subfigure}[b]{0.3\textwidth}
        \centering
        \includegraphics[width=\textwidth]{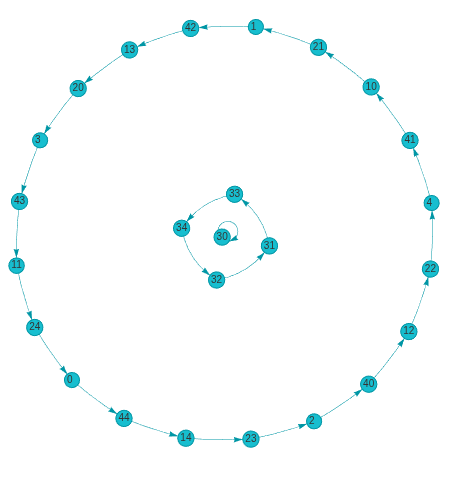}
        \caption{$n=2$}
        \label{fig:image1}
    \end{subfigure}
    \hfill
    \begin{subfigure}[b]{0.3\textwidth}
        \centering
        \includegraphics[width=\textwidth]{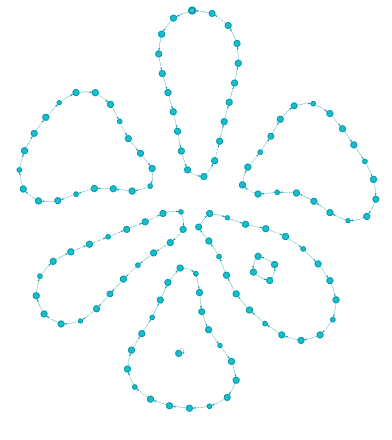}
        \caption{$n=3$}
        \label{fig:image2}
    \end{subfigure}
    \hfill
    \begin{subfigure}[b]{0.3\textwidth}
        \centering
        \includegraphics[width=\textwidth]{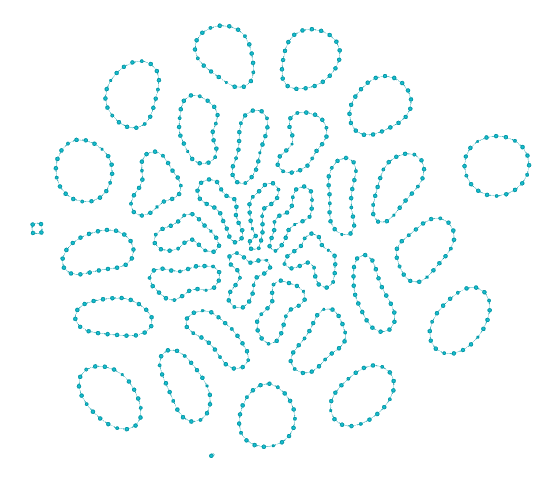}
        \caption{$n=4$}
        \label{fig:image3}
    \end{subfigure}
    \caption{Transition diagram of FDCA rule  $\langle0,0,0,0,2,3,0,4\rangle$ for the state $d=5$}
   \label{fig:0,0,0,0,2,3,0,4}
\end{figure}

\begin{figure}[ht]
    \centering
   
    \begin{subfigure}[b]{0.3\textwidth}
        \centering
        \includegraphics[width=\textwidth]{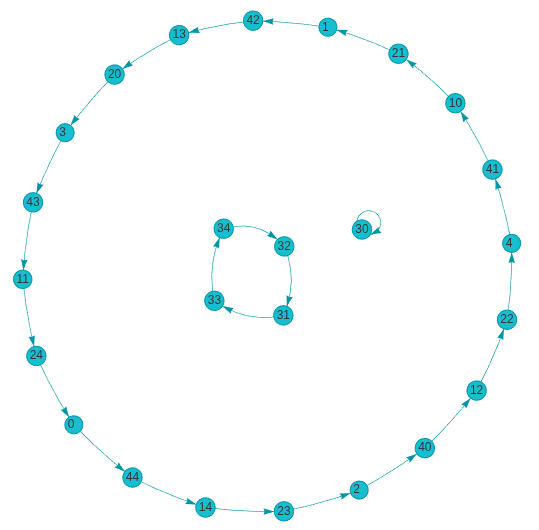}
        \caption{$n=2$}
        \label{fig:image1}
    \end{subfigure}
    \hfill
    \begin{subfigure}[b]{0.3\textwidth}
        \centering
        \includegraphics[width=\textwidth]{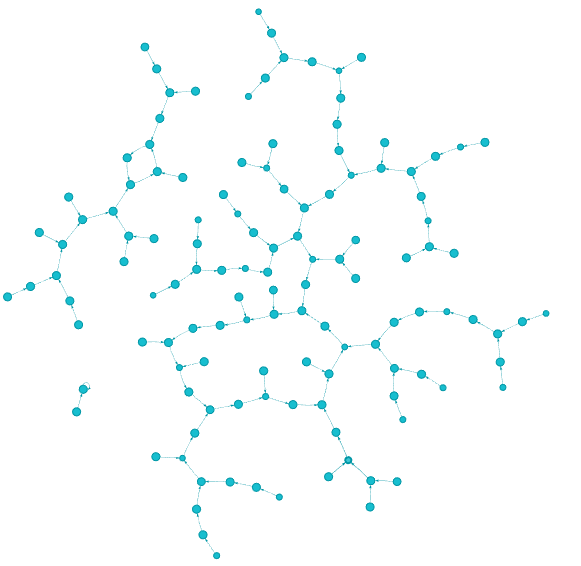}
        \caption{ $n=3$}
        \label{fig:image2}
    \end{subfigure}
    \hfill
    \begin{subfigure}[b]{0.3\textwidth}
        \centering
        \includegraphics[width=\textwidth]{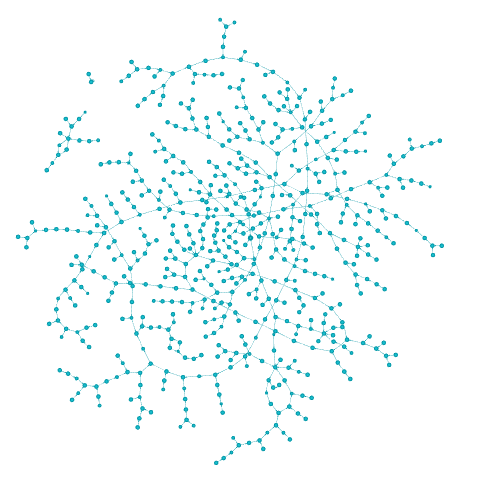}
        \caption{$n=4$}
        \label{fig:image3}
    \end{subfigure}
    \caption{Transition diagram of FDCA rule  $\langle2,0,0,0,2,3,0,4\rangle$ for the state $d=5$ }
     \label{fig:2,0,0,0,2,3,0,4}
\end{figure}

Using a graphical verification procedure, we can check whether the FDCA rules that satisfy the conditions are reversible or not. Let us consider an example to check the reversibility of the FDCA rule. As per Algorithm \ref{Algorithm_Rule_analysis}, the FDCA rule $\langle0,0,0,0,2,3,0,4\rangle$ is reversible for cells when $d=5$ because it satisfies all the reversibility conditions, whereas the FDCA rule $\langle2,0,0,0,2,3,0,4\rangle$ is not reversible (irreversible for some cells) since it violates the condition for coefficient $c_0$ for state $d=5$.
In order to verify visually, we plot the transition diagrams for some $n$ and analyze the state transitions. As shown in Figure \ref{fig:0,0,0,0,2,3,0,4}, when the state $d=5$, the FDCA rule $\langle0,0,0,0,2,3,0,4\rangle$ is reversible for $n=2,3,4, \cdots$. However, the FDCA rule $\langle2,0,0,0,2,3,0,4\rangle$ shown in Figure \ref{fig:2,0,0,0,2,3,0,4} is irreversible when $n=3,5$.

\vspace{-.3 cm}

\section{Conclusion}

In this work, we have done an algebraic characterization of the reversible rule to identify the necessary and sufficient conditions for reversibility of FDCAs. By using these conditions, one can synthesize all $d$-state FDCAs which are reversible for every lattice size under null boundary condition. Similarly, we can verify the reversibility of any FDCA rule by using a constant-time algorithm. In future work, this study can be extended to semi-reversibility of FDCA rules by analyzing cases where one or more of the identified conditions are violated. Also, finding such reversibility conditions for any $d$-state CAs will be another avenue of future research.

 \bibliographystyle{elsarticle-num} 
 \bibliography{reference.bib}

\end{document}